\documentclass[10pt]{article}
\usepackage{graphicx}

\def\Title#1{\begin{center} {\Large #1 } \end{center}}
\def\Author#1{\begin{center}{ \sc #1} \end{center}}
\def\Address#1{\begin{center}{ \it #1} \end{center}}

\newcommand\pubblock{\rightline{\begin{tabular}{l} Proceedings of the Fifth Annual LHCP\\ \pubnumber\\
         \pubdate  \end{tabular}}}

\newenvironment{Abstract}{\begin{quotation} \begin{center} 
             \large ABSTRACT \end{center}\bigskip 
      \begin{center}\begin{large}}{\end{large}\end{center} \end{quotation}}

\newenvironment{Presented}{\begin{quotation} \begin{center} 
             PRESENTED AT\end{center}\bigskip 
      \begin{center}\begin{large}}{\end{large}\end{center} \end{quotation}}

\def\Acknowledgements{\bigskip  \bigskip \begin{center} \begin{large}
             \bf ACKNOWLEDGEMENTS \end{large}\end{center}}

\def\beq{\begin{equation}}
\def\eeq#1{\label{#1}\end{equation}}
\def\eeqn{\end{equation}}

\def\beqa{\begin{eqnarray}}
\def\eeqa#1{\label{#1}\end{eqnarray}}
\def\eeqan{\end{eqnarray}}

\let\bar=\overbar

\def\Dslash{\not{\hbox{\kern-4pt $D$}}}
\def\dslash{\not{\hbox{\kern-2pt $\del$}}}

\def\msb{{\bar{\ssstyle M \kern -1pt S}}}

\usepackage{color}
\usepackage{wrapfig}

\newcommand\pubnumber{ }

\newcommand\pubdate{\today}

\def\affiliation{
On behalf of the CMS Collaboration, \\
Department of Physics and Astronomy \\
Seoul National University, Seoul, 08826, Republic of Korea}

\begin{document}

\large
\begin{titlepage}
\pubblock

\vfill
\Title{ Single Boson Production Cross Section Measurements in CMS  }
\vfill

\Author{ Kyeongpil Lee }
\Address{\affiliation}
\vfill
\begin{Abstract}

 Measurements of single boson production cross sections are presented. 
 They are based on proton-proton collision data at 8 and 13 TeV recorded with the CMS detector at the LHC. 
 Inclusive and differential cross sections with respect to various observables are measured in various phase spaces.
 These measurements are compared to perturbative QCD predictions and generally show good agreement with the prediction.

\end{Abstract}
\vfill

\begin{Presented}
The Fifth Annual Conference\\
 on Large Hadron Collider Physics \\
Shanghai Jiao Tong University, Shanghai, China\\ 
May 15-20, 2017
\end{Presented}
\vfill
\end{titlepage}
\def\thefootnote{\fnsymbol{footnote}}
\setcounter{footnote}{0}
%

\normalsize 


\section{Introduction}

Production of W and Z boson is one of the most important physics process at hadron colliders to understand Standard Model (SM) precisely.
This process is theoretically well known and the predictions are available at next-to-next-to-leading order (NNLO) in perturbative quantum chromodynamics (QCD).
In experimental point of view, high experimental precision can be obtained for weak boson measurement thanks to the clean signature from their leptonic decay channel with relatively large cross sections.
Therefore, measurements of weak boson cross sections provide many opportunities for the precise test of SM and the results can be used to constraint parton distribution functions (PDFs).
This article reports the production and differential cross section measurements of W, Z and Drell-Yan (DY) process using the Run 1 and 2 data collected with the CMS detector at LHC \cite{bib:CMS}.

\section{DY cross sections}
\subsection{Differential cross sections at 8 TeV}
Measurement of DY single and double differential cross sections at $\sqrt{s}=8$ TeV using the data corresponding to an integrated luminosity 19.7 fb$^{-1}$ is performed in \cite{bib:DY8TeV}.
Single differential cross sections are measured with respect to the dilepton mass $(d\sigma/dm)$ from 15 to 2 TeV divided by 41 bins,
and double differential cross sections are measured with respect to absolute dilepton rapidity up to 2.4 in 6 different dilepton mass range $(d^{2}\sigma/dmd|y|)$ from 20 GeV to 1.5 TeV with total 132 bins.

Based on the data passing double lepton trigger, event selection requires two isolated leptons with $P_{T}^{lead} > 20$ GeV and $P_{T}^{sub} > 10$ GeV for both electron and muon channel.
After subtracting the backgrounds estimated by data-driven method, unfolding correction is applied to take into account the bin migration from the detector resolution and FSR effect.
Acceptance and efficiency corrections are also applied.

\begin{figure}[htb]
\centering
\includegraphics[height=2.5in]{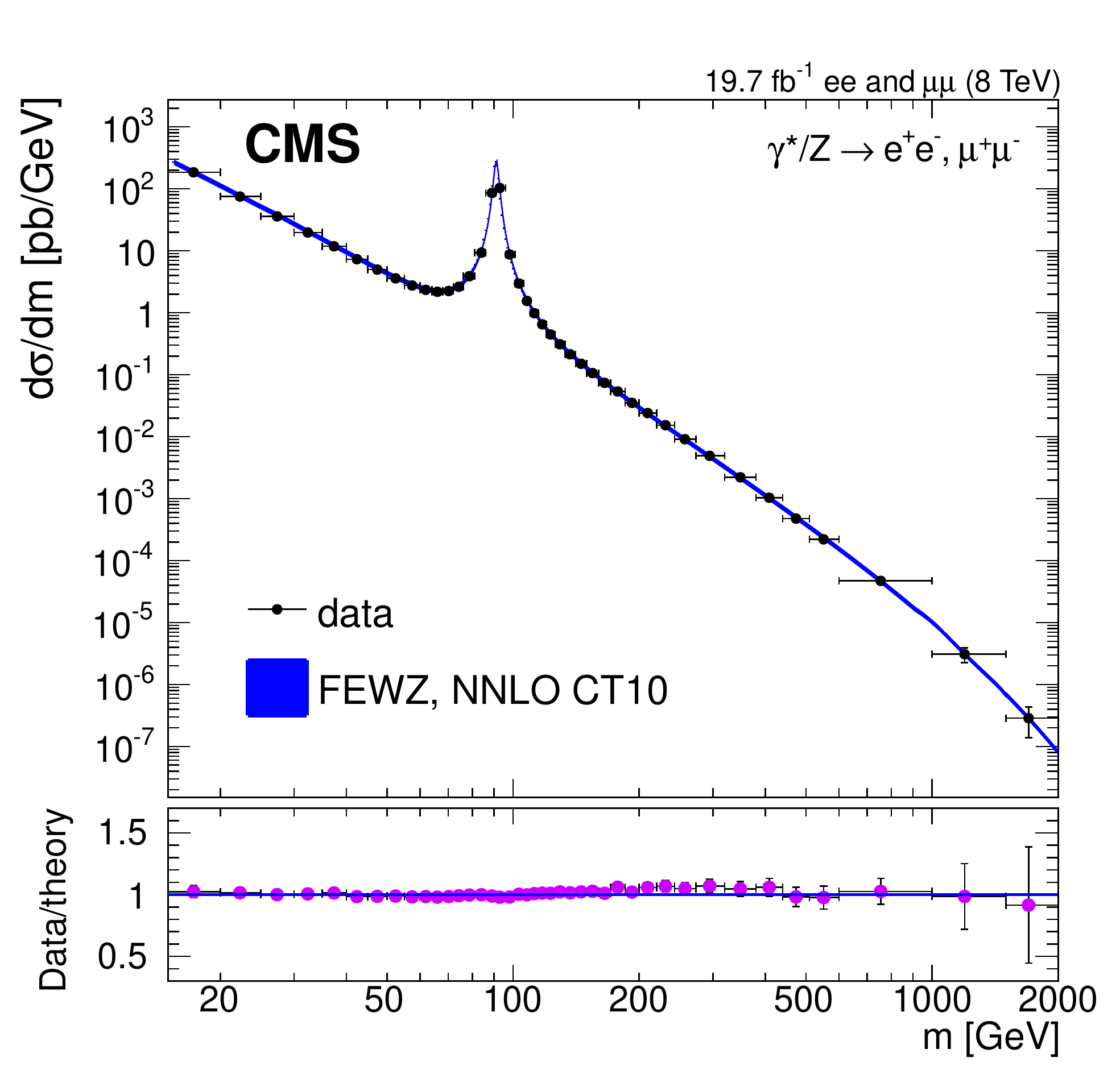}
\includegraphics[height=2.4in]{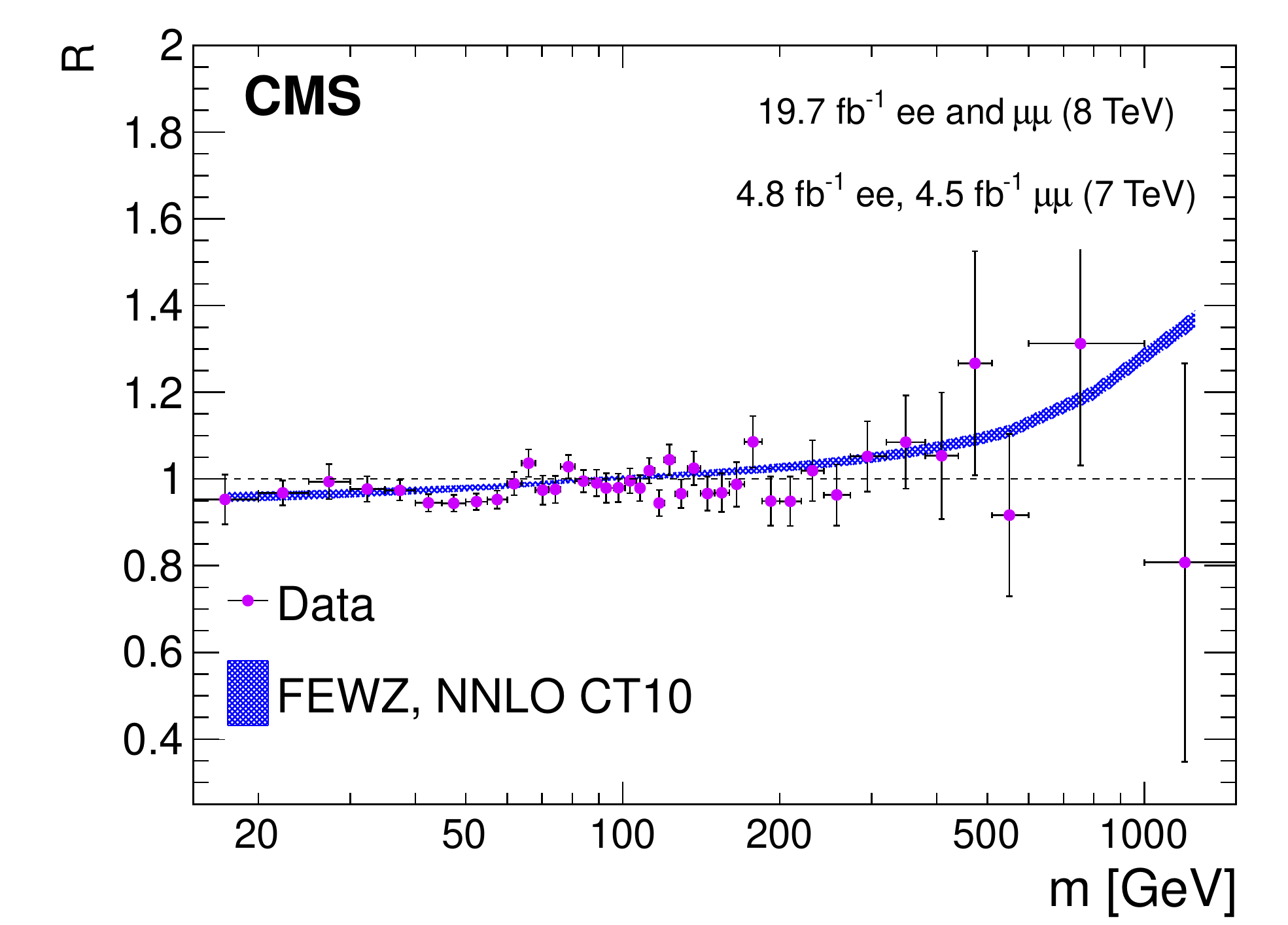}
\caption{ Left: Result of $d\sigma/dm$ after combination of muon and electron channel compared with the theoretical prediction from FEWZ with CT10 PDF set. 
Right: Result of double ratio of $d\sigma/dm$ between 7 and 8 TeV with the theoretical prediction \cite{bib:DY8TeV}. }
\label{fig:DY8TeV_XSec_1D}
\end{figure}

\begin{figure}[htb]
\centering
\includegraphics[height=2in]{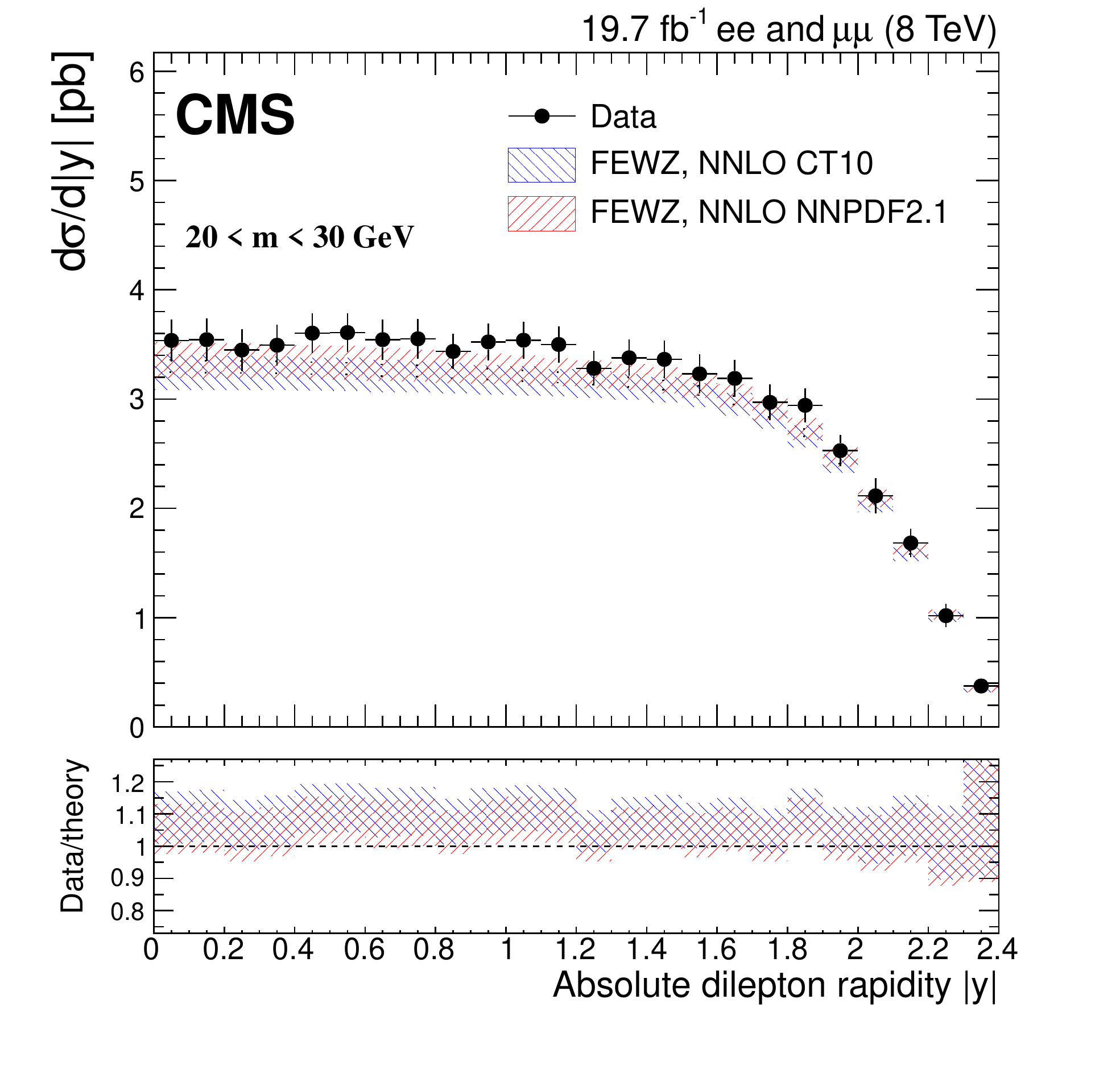}
\includegraphics[height=2in]{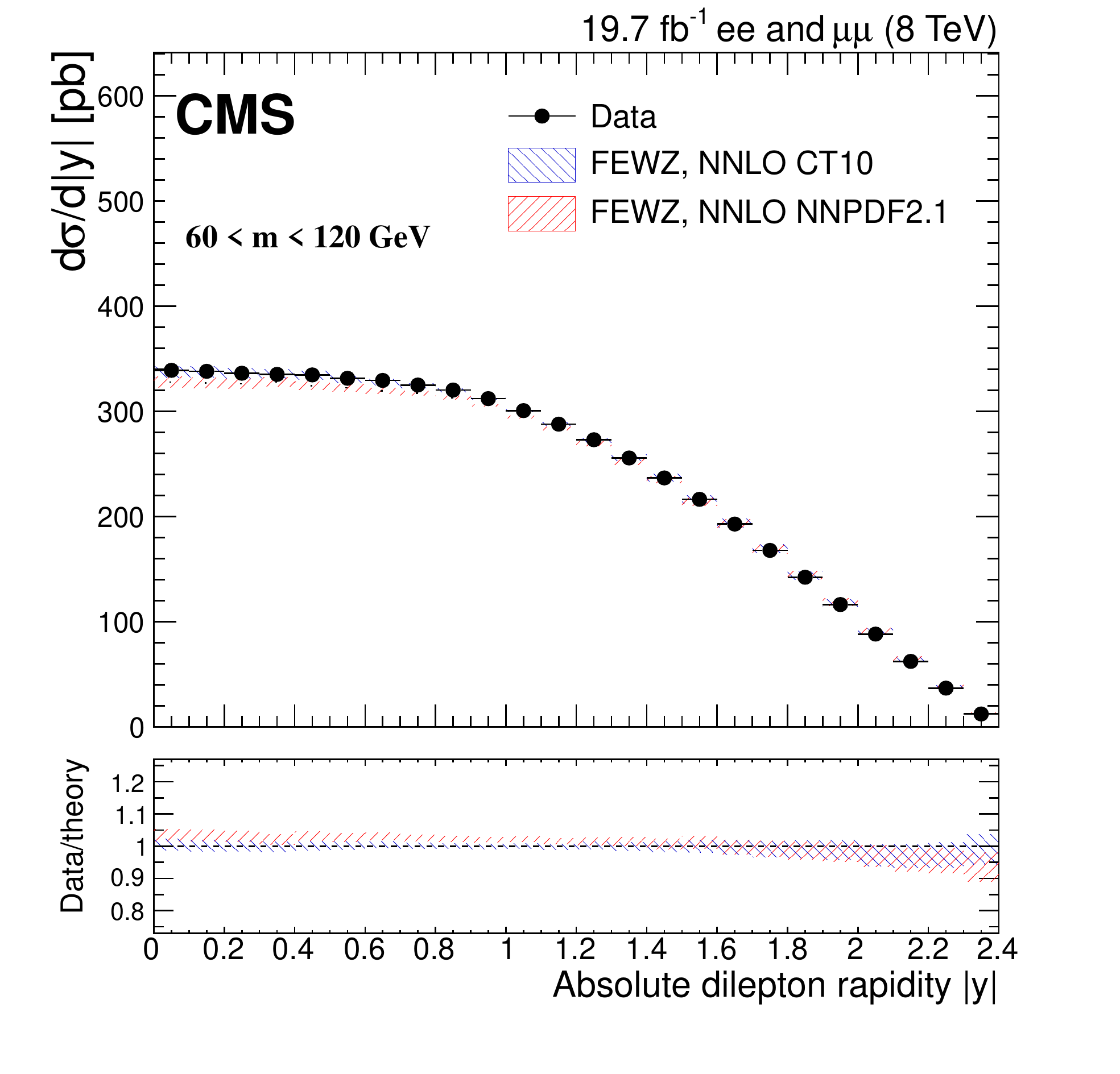}
\includegraphics[height=2in]{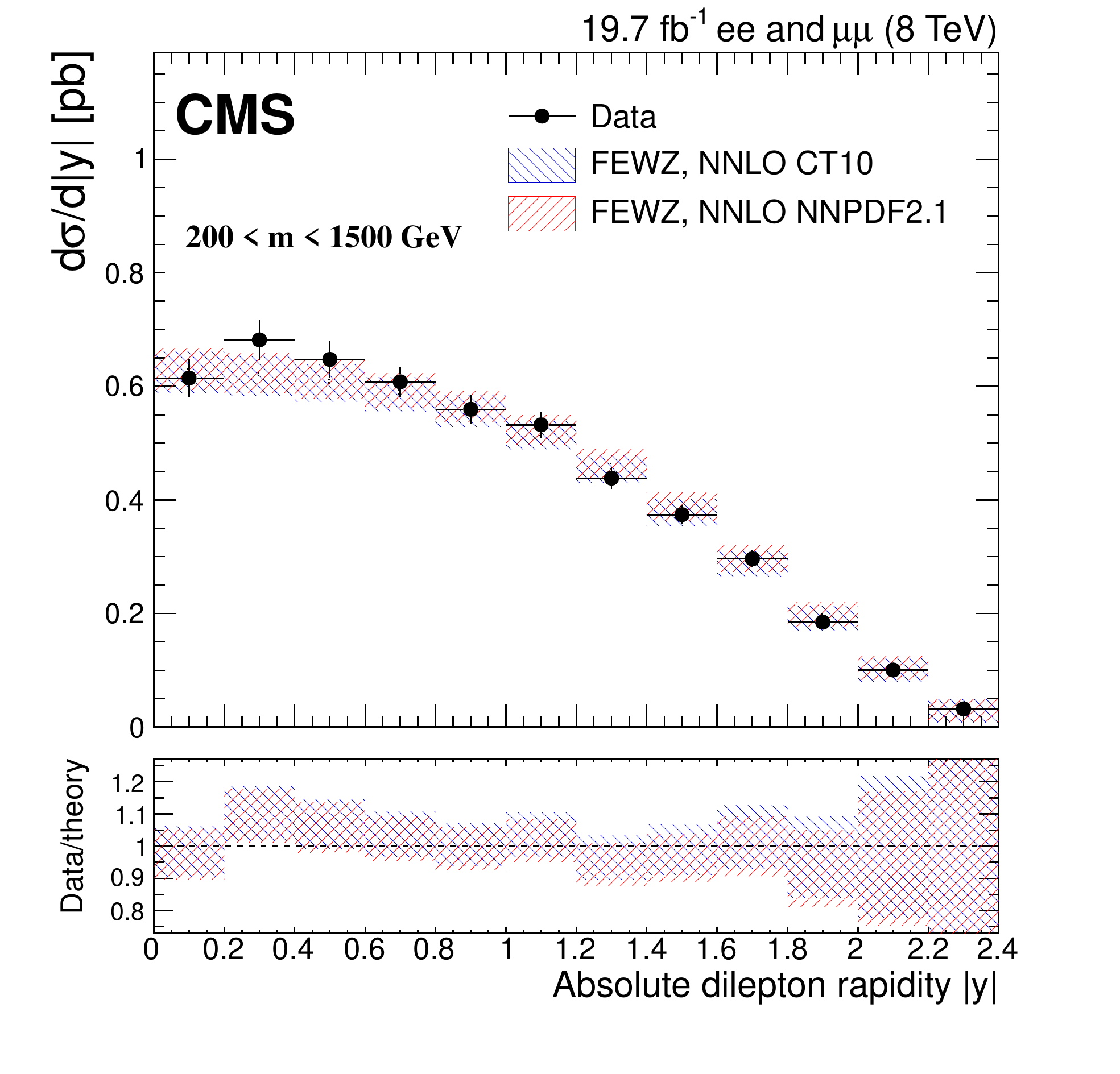}
\caption{ Differential cross section with respect to dilepton rapidity $d\sigma/d|y|$ in the low mass, Z resonance and high mass region respectively.
They are compared with the theoretical predictions from FEWZ with NNPDF 2.1 and CT10 PDF sets \cite{bib:DY8TeV}.  }
\label{fig:DY8TeV_XSec_2D}
\end{figure}

The left side of Fig.~\ref{fig:DY8TeV_XSec_1D} is the result of $d\sigma/dm$ after the combination between electron and muon channel for greater precision.
Experimental result is compared with the theoretical prediction calculated by FEWZ with CT10 PDF set, which shows good agreement with the data within a few \% over entire mass range.
Double ratio, which is the ratio of normalized differential cross section divided by Z cross section between two different center of mass energies (7 and 8 TeV), is shown in the right plot.
Double ratio has higher sensitivity to NNLO QCD effects and can give precise constraint on the PDFs.
It also generally shows good agreement with the theoretical prediction from FEWZ with CT10 PDF set.
$d^{2}\sigma/dmd|y|$ results for low mass, Z resonance and high mass region are shown in Fig.~\ref{fig:DY8TeV_XSec_2D}. 
Acceptance correction is not applied to double differential cross sections to minimize the theoretical uncertainties.
Experimental results are compared to the theoretical predictions with NNPDF 2.1 and CT10 PDF sets in NNLO calculation, and data and theory are generally in good agreement.
The other results in different mass range and double ratio of $d^{2}\sigma/dmd|y|$ are available in \cite{bib:DY8TeV}.

Another analysis \cite{bib:DYPhiStar8TeV} based on the same amount of the data with \cite{bib:DY8TeV} measures the differential cross sections with respect to the $\phi^{*}$ variable,
which depends only on lepton angles but is correlated to the transverse momentum of dilepton system $(q_{T})$. 
Therefore, this measurement can give the opportunity to probe even low $q_{T}$ region with significantly better resolution than $q_{T}$.
In the event selection procedure, $P_{T}$ of leading (sub-leading) lepton should be larger than 30 (20) GeV with $|\eta|$ less than 2.1 (2.4).
Also dilepton mass is confined within Z resonance region from 60 GeV to 120 GeV to minimize the background contamination.
After background subtraction by cut and count method, detector resolution effect is taken into account by unfolding correction.

\begin{figure}[htb]
\centering
\includegraphics[height=2.45in]{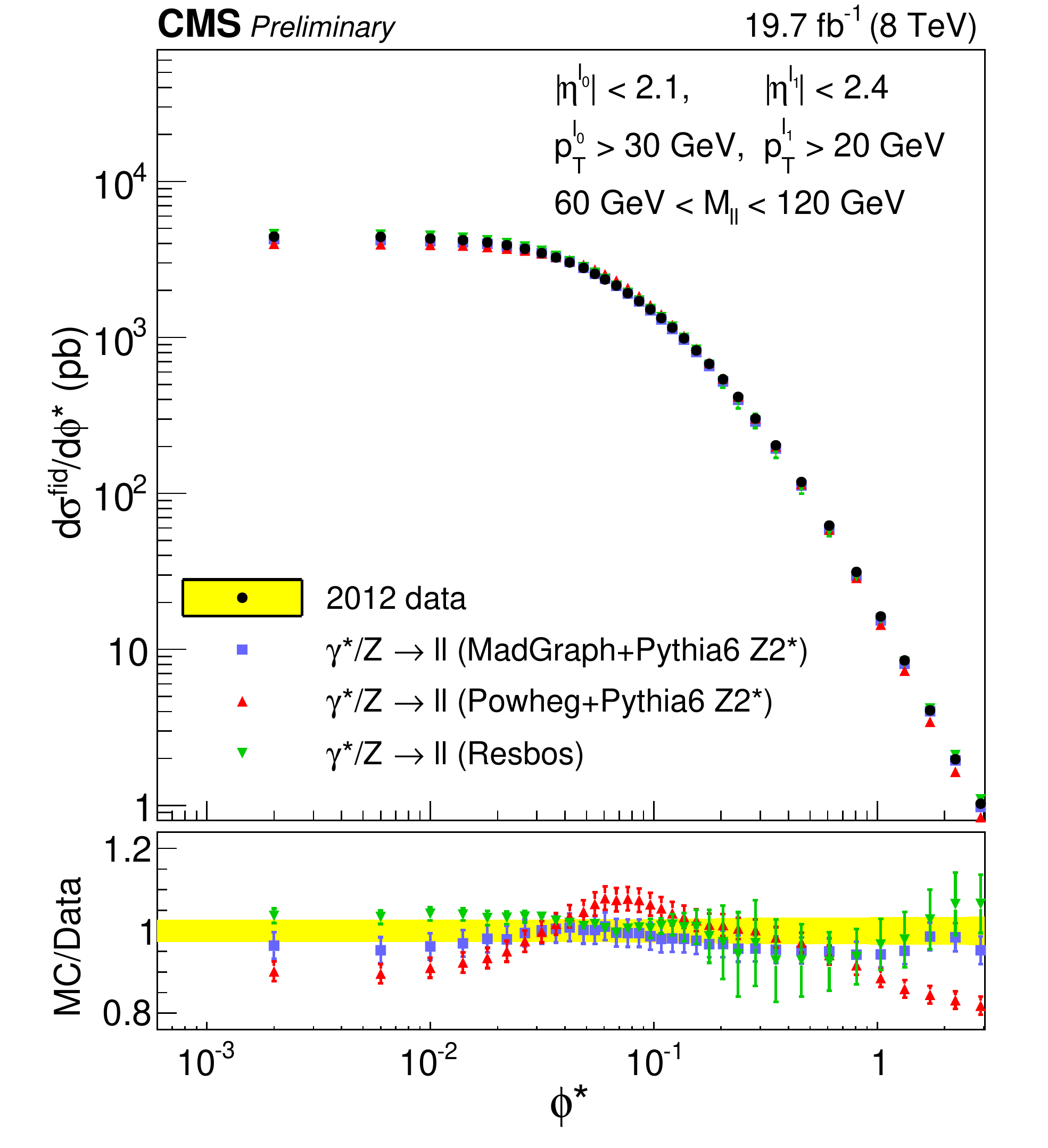}
\includegraphics[height=2.45in]{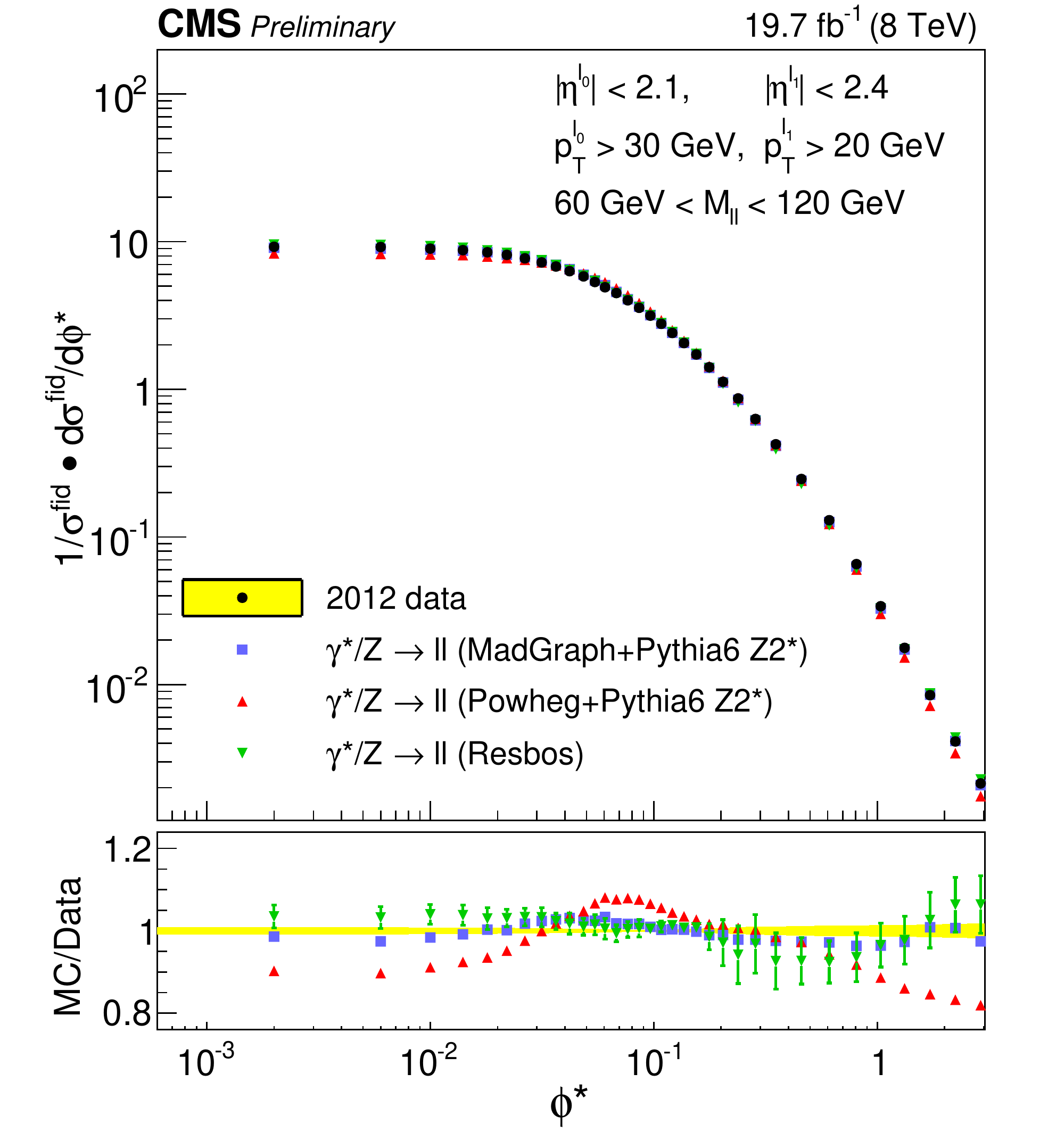}
\caption{ Left: Differential cross sections of $d\sigma/d\phi^{*}$ with the theoretical predictions from Madgraph, Powheg and Resbos. 
Right: Normalized differential cross sections by total fiducial cross section \cite{bib:DYPhiStar8TeV}.}
\label{fig:DYPhiStar8TeV_XSec}
\end{figure}

Fig.~\ref{fig:DYPhiStar8TeV_XSec} shows the fiducial and normalized differential cross sections after combination between muon and electron channel. 
Especially in the normalized cross sections, total experimental uncertainty is about 1-2\% level due to the high precision of lepton angular variables.
It is also compared to the various theoretical predictions from Madgraph, Powheg and Resbos, and each prediction shows different behavior.
However, none of them describes well the data distribution over entire $\phi^{*}$ range.

\subsection{Differential cross sections at 13 TeV}

Single differential cross sections $d\sigma/dm$ of DY process are also measured using 13 TeV data corresponding to an integrated luminosity 2.8 fb$^{-1}$ with dimuon channel \cite{bib:DY13TeV}.
Analysis strategy is similar with the one for 8 TeV data \cite{bib:DY8TeV}, 
but the mass range is extended up to 3 TeV with more bins thanks to the increase of center of mass energy, 
and leading $P_{T}$ threshold slightly increases to 22 GeV because of trigger requirement.

\begin{figure}[htb]
\centering
\includegraphics[height=2.3in]{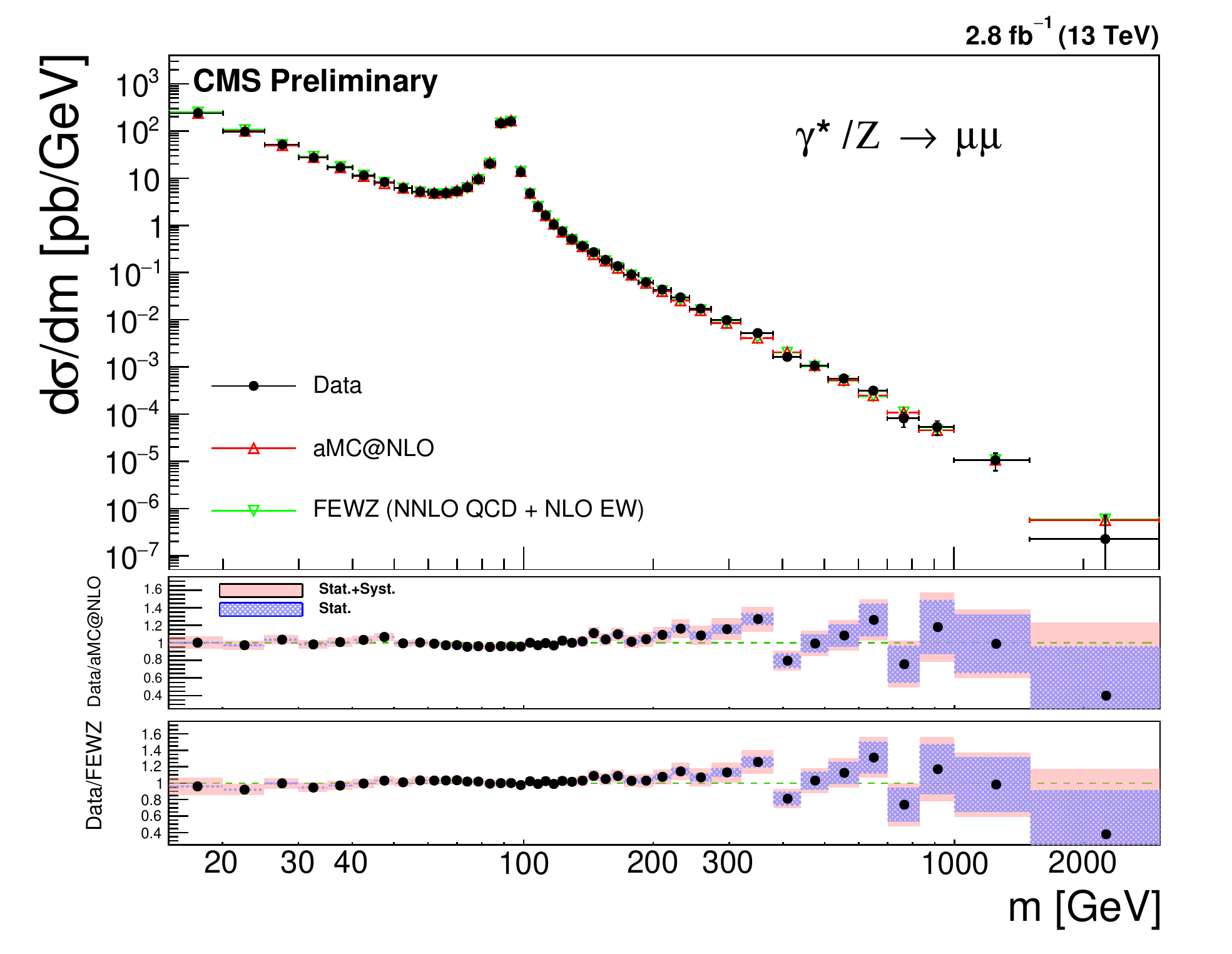}
\includegraphics[height=2.3in]{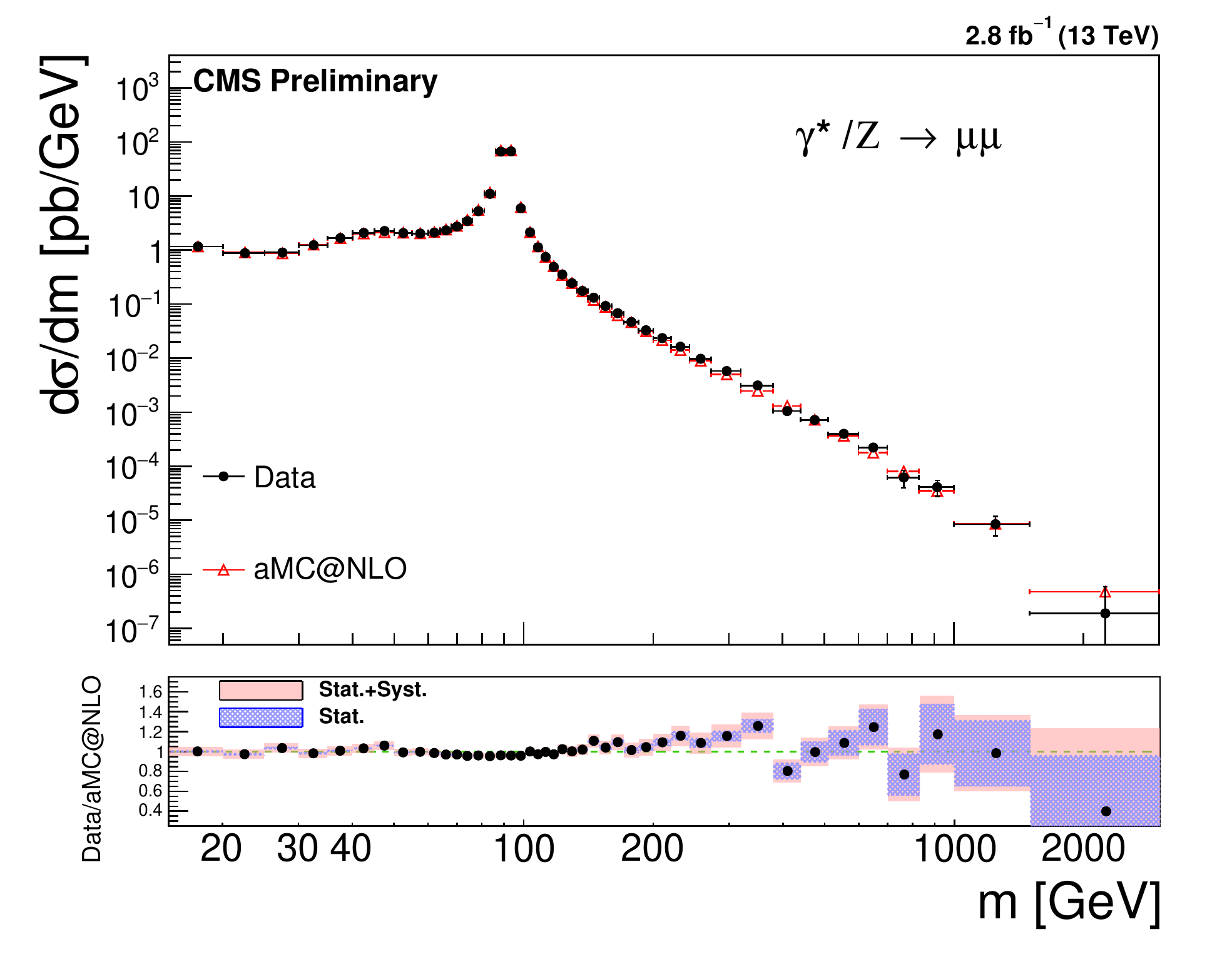}
\caption{ Left: FSR-corrected $d\sigma/dm$ result in full phase space.
Right: $d\sigma/dm$ result in fiducial region without FSR correction.
Both results are dimuon channel only \cite{bib:DY13TeV}. }
\label{fig:DY13TeV_XSec_1D}
\end{figure}

Fig.~\ref{fig:DY13TeV_XSec_1D} shows the results of differential cross sections. 
Left one corresponds to the cross sections in full phase space with FSR correction by unfolding to dressed level.
It is compared to aMC@NLO and FEWZ prediction with NNPDF 3.0 PDF set.
The results on the right side is the fiducial cross section without FSR correction to minimize the theoretical inputs, 
and it is also compared to the theoretical predictions from aMC@NLO. 
Both results show the agreement between data and theory.

\section{W and Z cross sections}
\subsection{Inclusive W and Z cross sections at 13 TeV}
Measurement of inclusive W and Z cross sections is performed using the first 13 TeV data in Run 2 of LHC, which corresponds to an integrated luminosity 43 pb$^{-1}$ \cite{bib:WZ13TeV}.
In order to select Z candidate, symmetric $P_{T}$ cut (25 GeV) is applied for both leptons and their invariant mass should be within the mass range from 60 to 120 GeV.
The backgrounds, which have less than 1\% contribution in Z resonance region, are estimated by Monte-Carlo (MC) predictions that simulates $\tau$ decay of Z boson, top-quark pair production and dibosons.
On the other hand, signal extraction of W candidate is done by fitting $E^{miss}_{T}$ distribution using W signal, EWK and QCD background shape.

\begin{figure}[htb]
\centering
\includegraphics[height=2.2in]{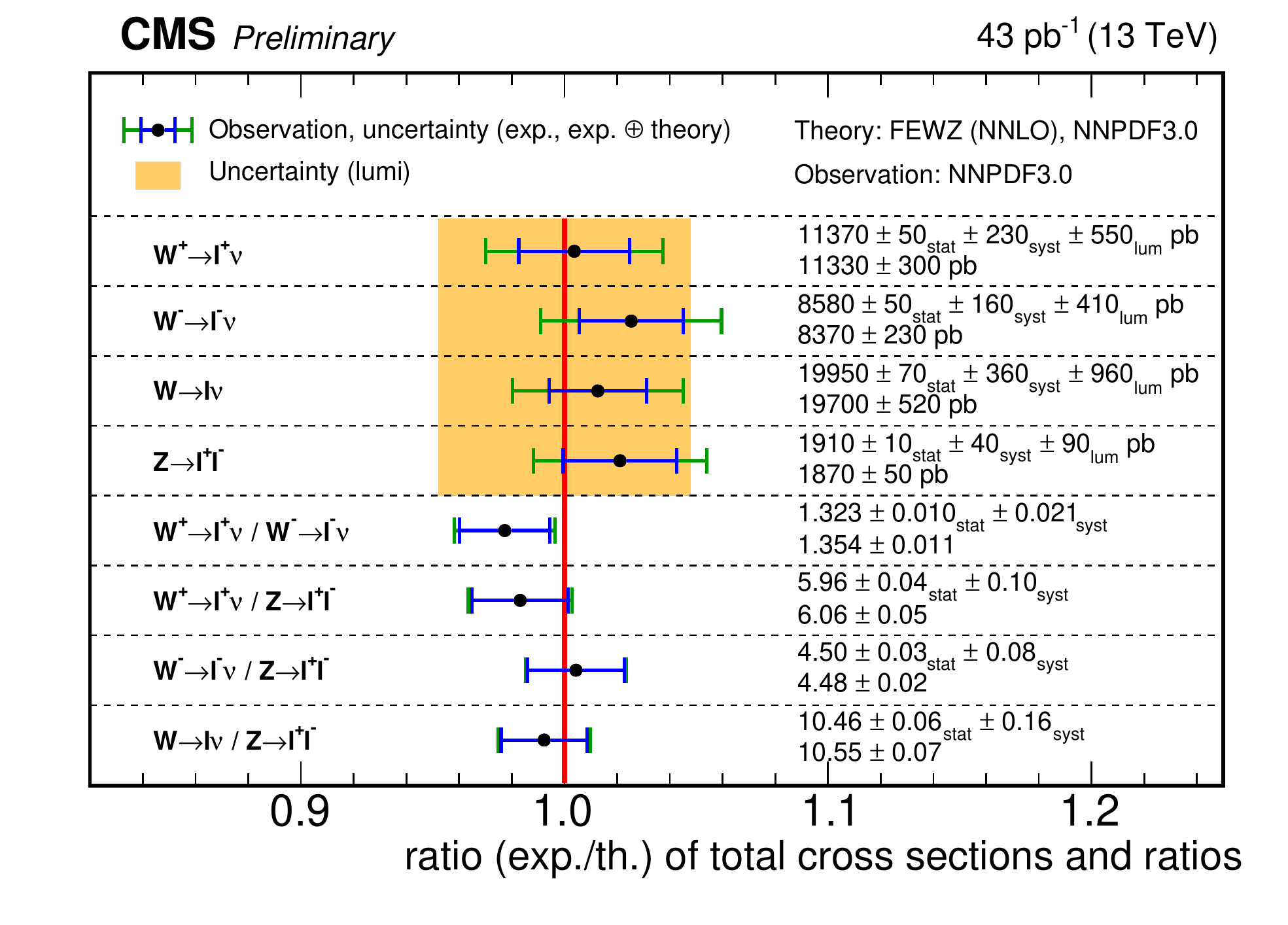}
\includegraphics[height=2.2in]{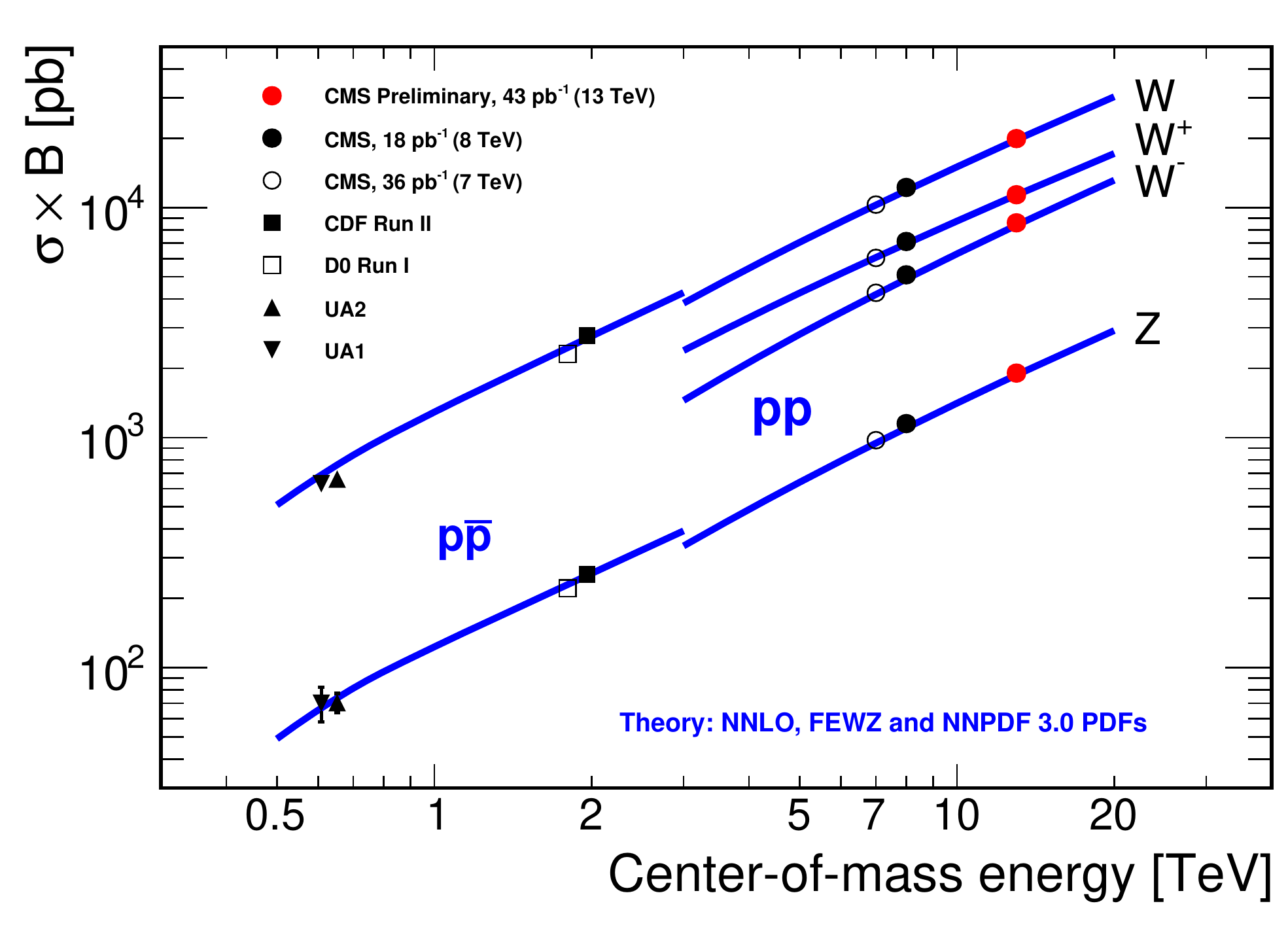}
\caption{ Left: Summary of ratios between experimental result and theoretical prediction of W and Z cross sections and their ratios.
Right:  W and Z cross section results from various experiments with different center of mass energies.
Both results are compared with the theoretical prediction at NNLO calculated by FEWZ with NNPDF 3.0 PDF set \cite{bib:WZ13TeV}. }
\label{fig:WZ13TeV_XSec}
\end{figure}

The results of cross sections are summarized in Fig.~\ref{fig:WZ13TeV_XSec}.
According to the left plot, measured cross sections are slightly higher than the theory prediction but generally they agree well within uncertainties.
Also, right plot shows that W and Z cross sections from various measurements performed so far have good agreement with theory in each center of mass energy.

\subsection{Z differential cross sections at 13 TeV}
CMS collaboration has also performed differential cross section measurement for Z boson using full 2015 data which corresponds to an integrated luminosity 2.8 fb$^{-1}$ in dimuon channel \cite{bib:ZDiff13TeV}.
This measurement presents various differential cross sections with respect to dilepton transverse momentum $P_{T}^{\mu^{+}\mu^{-}}$, $\phi^{*}$, absolute dilepton rapidity $|y|$, and single lepton transverse momentum $P_{T}^{\mu}$.
Analysis procedure is similar with the one for inclusive measurement \cite{bib:WZ13TeV}, 
but unfolding correction is applied in this analysis to take into account the bin migration effect from the detector resolution.

\begin{figure}[htb]
\centering
\includegraphics[height=2in]{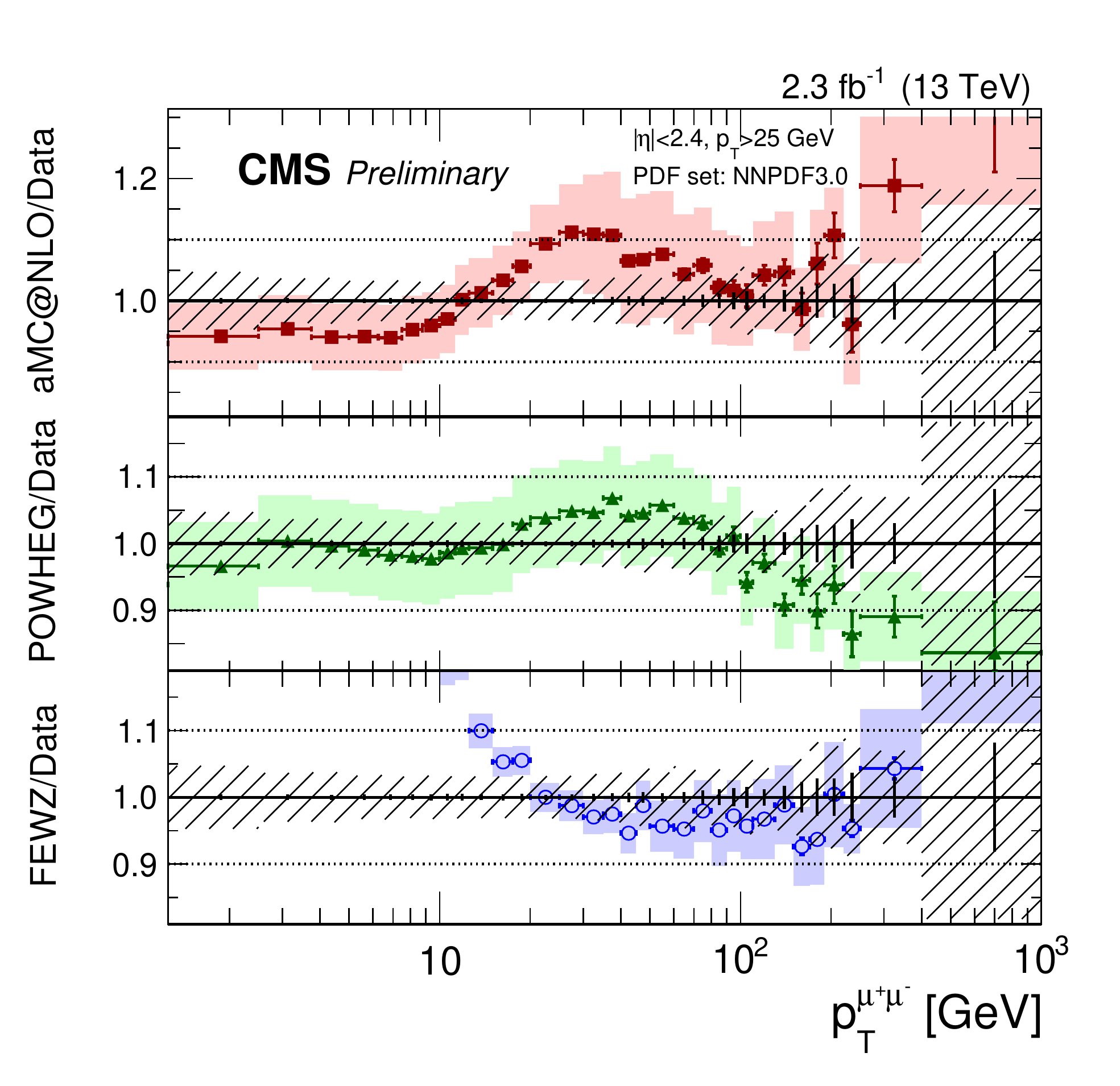}
\includegraphics[height=2in]{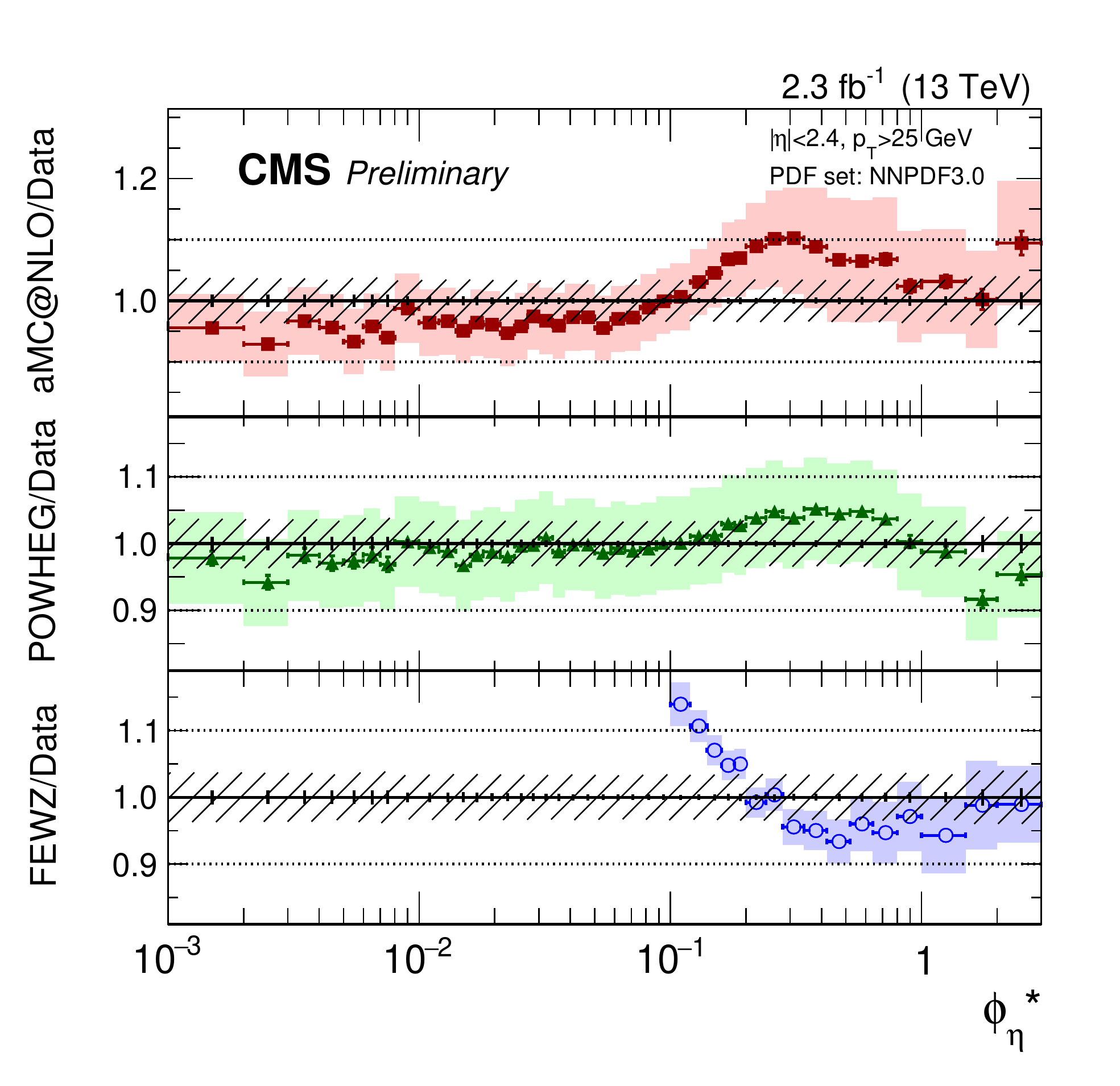}
\includegraphics[height=2in]{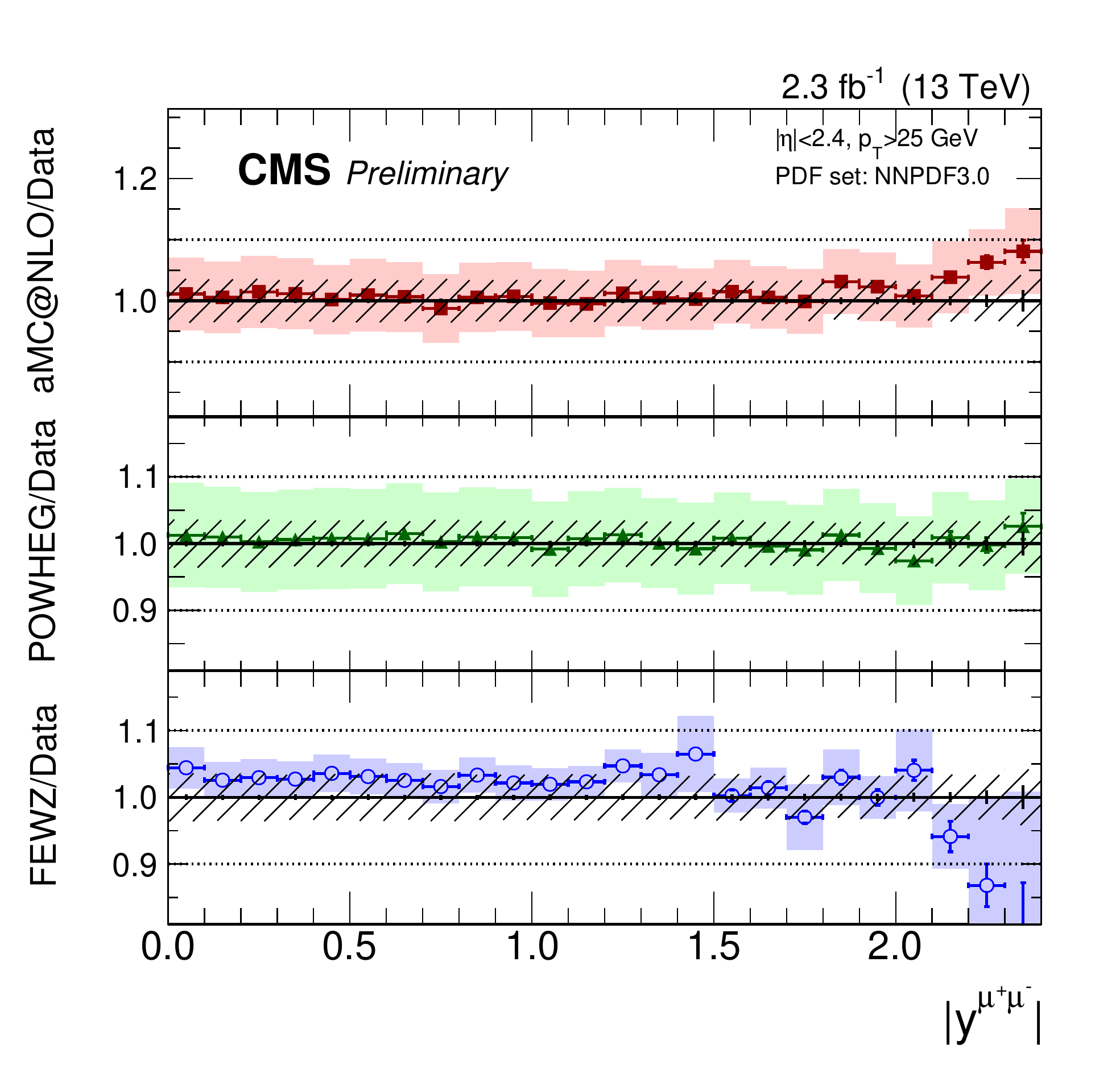}
\caption{ Ratios of differential cross sections with respect to $P_{T}^{\mu^{+}\mu^{-}}$, $\phi^{*}$ and $|y|$ between experimental results and theory predictions from aMC@NLO, Powheg and FEWZ at NNLO \cite{bib:ZDiff13TeV}.  }
\label{fig:ZDiff13TeV_XSecRatio}
\end{figure}

Ratios of differential cross section results between data and theory are shown in Fig.~\ref{fig:ZDiff13TeV_XSecRatio} with various theoretical predictions from aMC@NLO, Powheg and FEWZ at NNLO.
In $d\sigma/dP_{T}^{\mu^{+}\mu^{-}}$ result, aMC@NLO and Powheg agree well with the data in low $P_{T}^{\mu^{+}\mu^{-}}$ region, 
but they have opposite behavior as $P_{T}^{\mu^{+}\mu^{-}}$ becomes higher. On the other hand, FEWZ has large deviation in the low $P_{T}^{\mu^{+}\mu^{-}}$ region, which comes from the absence of resummation. This deviation also presents in low $\phi^{*}$ region. In the case of $d\sigma/d|y|$, all theory predictions describe the data well over entire range.

\begin{wrapfigure}[15]{R}{0.4\textwidth}
\centering
\vspace{-15pt}
\includegraphics[height=2.3in, width=2.5in]{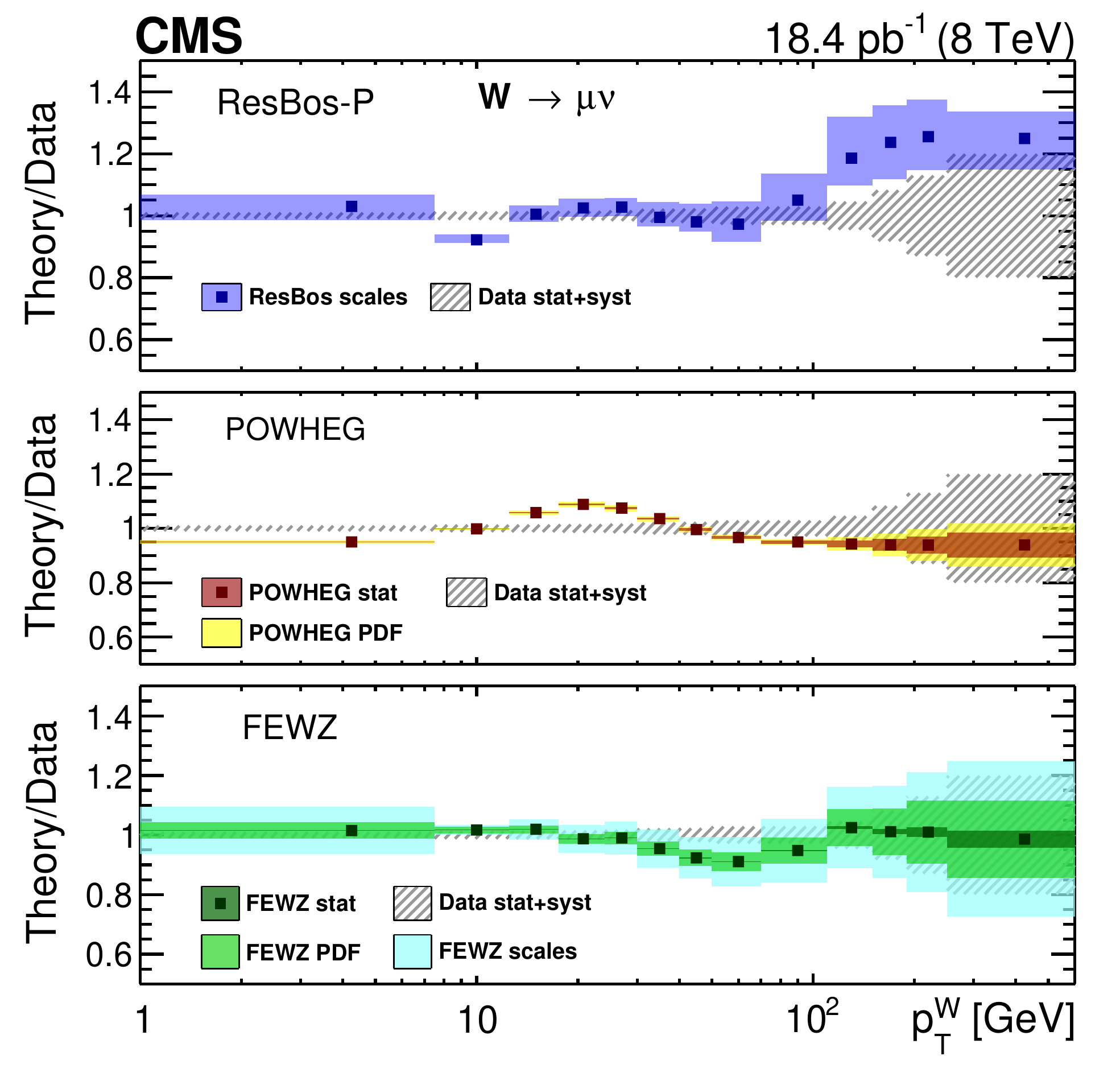}
\caption{W differential cross sections with respect to W $P_{T}$ ($d\sigma/dP_{T}^{W}$) \cite{bib:WZDiff8TeV}.}
\vspace{15pt}
\label{fig:WZDiff8TeV_XSecRatio}
\end{wrapfigure}

\subsection{W and Z differential cross sections and their ratios at 8 TeV}
Differential cross sections for W are measured using the 8 TeV data as well as Z \cite{bib:WZDiff8TeV}.
The cross section ratios between W$^{+}$ and W$^{-}$, and Z and W are also included.
This analysis is based on a special dataset with low luminosity at 8 TeV with an integrated luminosity about 18 pb$^{-1}$.
This dataset has small number of collisions per bunch-crossing (low pile-up), which leads to small backgrounds and improved resolution.
For the event selection, $P_{T} > 25 (20)$ GeV and $|\eta| <$ 2.4 (2.1) are required for electron (muon) channel.
Signal extraction for W and Z is similar with the other analyses in previous sections.

Fig.~\ref{fig:WZDiff8TeV_XSecRatio} shows the ratio of the W differential cross sections between data and theory prediction from Resbos, Powheg and FEWZ.
According to the plot, Resbos under-predicts the data near 10 GeV, and overestimates above 100 GeV by about 20\%. 
On the other hand, Powheg and FEWZ have about 10\% deviation near 25 GeV and 60 GeV in opposite direction.

\vspace{5mm}



\section{Conclusions}
Measurements for single boson production and differential cross sections are presented especially for W and Z in CMS, which provide the precise test of SM.
Using both Run 1 and Run 2 data with different center of mass energies, many kinds of inclusive and differential cross sections are measured,
and all of them generally show good agreement with the theoretical predictions within the uncertainties.

\Acknowledgements
K. Lee is supported in part by the National Research Foundation of Korea (NRF)
funded by the Korea government (NRF-2015R1C1A1A01053087).

\end{document}